\documentclass[prb,superscriptaddress,twocolumn]{revtex4-1}

\usepackage{amsmath}
\usepackage{amssymb}
\usepackage{graphicx}
\usepackage{epsfig}
\usepackage{dcolumn}
\usepackage{textcomp}
\usepackage{bm}
\usepackage[normalem]{ulem} 
\usepackage{color} 
\usepackage{subfigure}
\usepackage{pstricks}

\begin{document}

\title[Correlated transport in junction arrays: incoherent Cooper-pairs and hot electrons]{Correlated transport through junction arrays in the small Josephson energy limit: incoherent Cooper-pairs and hot electrons} 

\author{Jared H Cole}
\email{jared.cole@rmit.edu.au}
\address{%
Chemical and Quantum Physics, School of Applied Sciences, RMIT University, Melbourne, Victoria 3001, Australia.
}
\address{Institute f\"ur Theoretische Festk\"orperphysik, Karlsruhe Institute of Technology, 76128 Karlsruhe, Germany}
\author{Juha Lepp\"akangas}
\address{Institute f\"ur Theoretische Festk\"orperphysik, Karlsruhe Institute of Technology, 76128 Karlsruhe, Germany}
\address{Department of Microtechnology and Nanoscience, MC2, Chalmers University of Technology, SE-41296 G\"oteborg, Sweden}
\author{Michael Marthaler}
\address{Institute f\"ur Theoretische Festk\"orperphysik, Karlsruhe Institute of Technology, 76128 Karlsruhe, Germany}



\date{\today}
             
\begin{abstract}
We study correlated transport in a Josephson junction array for small Josephson energies. In this regime transport is dominated by Cooper-pair hopping, although we observe that quasiparticles can not be neglected. We assume that the energy dissipated by a Cooper-pair is absorbed by the intrinsic impedance of the array. This allows us to formulate explicit Cooper-pair hopping rates without adding any parameters to the system. We show that the current through the array is correlated and crucially, these correlations rely fundamentally on the interplay between the Cooper-pairs and equilibrium quasiparticles. 
\end{abstract}

\maketitle

\section{Introduction}
Linear arrays of Josephson junctions display surprisingly complex behaviour which belies their simple circuit diagrams.  The earliest work on these circuits considered the strong, long-range interaction between charges within the array\cite{Likharev:1989tg,Amman:1989tj,BenJacob:1989tt, Middleton:1993ww,Nazarov:1994wj}.  This Coulomb interaction between charges decays exponentially over a characteristic length that is set by the values of the circuit capacitances.  During conduction, the repulsive interaction is counteracted by the voltage applied across the circuit, which `pushes' the charges closer together.  The equilibrium charge configuration is then reached when these two energies are balanced, resulting in a periodic charge distribution across the array.  The movement of such a periodic charge state through the array in turn causes a periodic modulation of the current leaving the array.  Through this process the statistical correlations in the charge distribution are converted into correlations in the current, ie.\ correlated transport.  The charges within a Josephson junction array therefore display statistical correlations in both \emph{space} and \emph{time}, at the few electron level.
The resulting correlated transport makes them a candidate circuit for applications in quantum metrology and current standards\cite{Delsing:1992tq}.  In contrast to the state of the art in charge-pump metrology circuits\cite{Giblin:2012wi,Pekola:2012ti},  the use of Josephson junction arrays (JJA) in metrology is currently limited by the precision of both the theoretical models and the experimental devices.

Although qualitative agreement between semi-classical theory and experiment has been seen in several devices\cite{Haviland:1996wz,Haviland:2000tt,Andersson:2003ui,Agren:2000ux,Agren:2001vu}, there are still many open questions involving hysteretic and thermal effects\cite{ZorinAB:2006cv,Homfeld:2011fw}, the role of quasiparticles and charge noise.  The ultimate goal of a full quantum theory of these devices poses several challenges.  Obtaining quantitative agreement at the level now possible in simpler Josephson devices~\cite{Martinis:2005kw,Schuster:2007ki,Pashkin:2011ev,Cole:2010kw,Bushev:2010gw,OConnell:2010br} is difficult due to the interplay of charging energy, Josephson energy, charge noise and thermal effects.  In addition, most experiments have focused on current-voltage characteristics\cite{Haviland:1996wz, Delsing:1992tq} or microwave driven experiments~\cite{Delsing:1990tz,Delsing:1992tq} which limits the amount of information available for comparison.  An important exception are recent experiments\cite{Bylander:2005gr,Bylander:2007gb,Bylander:2005gc} where the correlated transport of charges within a JJA was measured directly using a radio-frequency single-electron transistor (rf-SET).  This provided detailed temporal information about the charge transport as well as the usual current-voltage characteristics.  Of particular note is the observation of correlated transport for a range of magnetic fields, suggesting a smooth transition from normal- to super-current\cite{Bylander:2007gb}.

Using a relatively simple model, we demonstrate that correlated transport can be carried by incoherent Cooper-pair hopping in the limit of small $E_J$.  We study the effect of equilibrium quasiparticles and incoherent Cooper-pair transport where the charge hopping is modelled using P(E) theory with the arrays \emph{own intrinsic impedance} (in contrast to Ref.~\onlinecite{Ho:2010ft} where an arbitrary, large impedance was assumed).  Considering the effect of both temperature and magnetic field on conduction, we also show that the experimentally observed response is consistent with some level of self-heating of the array.  This model opens the way forward for quantitative comparison between theory and future experiments and the development of a full microscopic model of transport in junction array circuits.  

In the following section we derive our noise model due to the intrinsic array impedance and discuss details of the simulation method.  In section 3 we consider the conduction through the array and show how correlated transport arises as a function of voltage.  We then study the regime where Cooper-pair and quasiparticle transport coexists and results in a novel conduction regime (sections 4 and 5).  Finally, in section 6 we consider the effect of magnetic field and the reduction of the superconducting gap on transport through the array.

\section{Noise model for long arrays in the small $E_J$ limit}

\begin{figure} [t!]
\centering{\includegraphics[width=0.9\columnwidth]{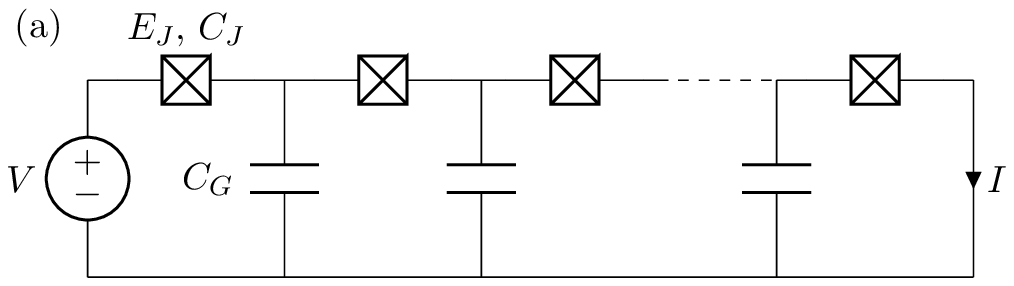}} \\
\vspace{0.3cm}
\centering{\includegraphics[width=0.9\columnwidth]{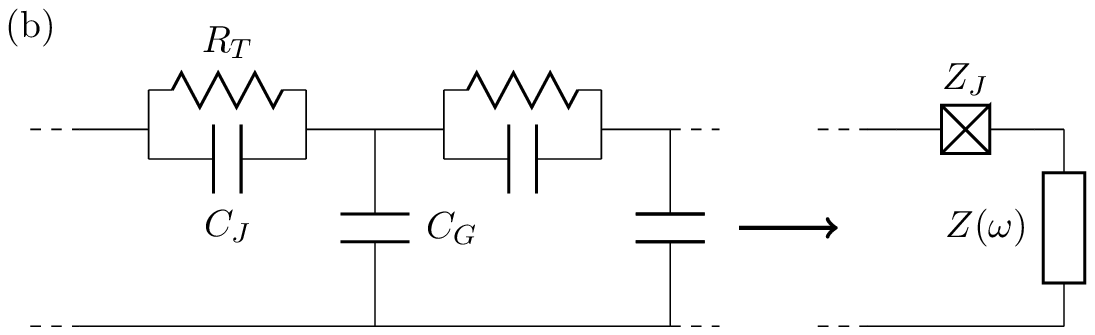}} \\
\vspace{0.3cm}
\centering{\includegraphics[width=0.9\columnwidth]{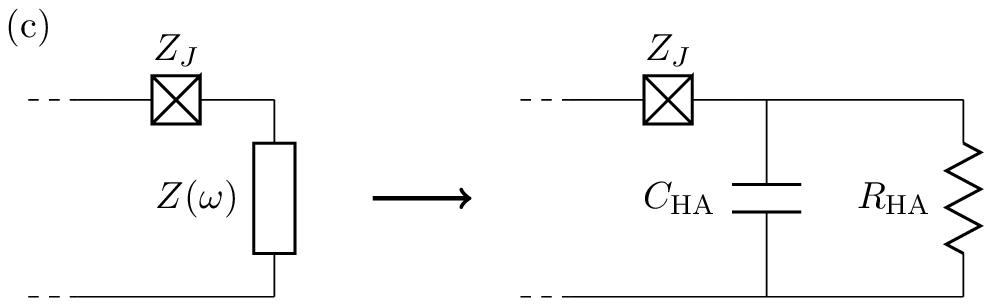}} 
\caption{(a) JJA circuit under consideration, consisting of a linear chain of Josephson junctions with Josephson energy $E_J$ and capacitance $C_J$.  The circuit is driven by a voltage source $V$ and each junction sees an effective capacitance to ground $C_G$. (b) We compute the effective impedance of the half-array seen by a particular Josephson junction, assuming that the rest of the array is composed of effective capacitors and the normal tunnel resistance $R_T$ for each junction.  (c) The half-array impedance is approximated assuming the capacitive and resistive components of the array decouple resulting in an effective parallel RC circuit. \label{fig:cctdiag}}
\end{figure}

Throughout this work, we consider a simple JJA circuit with a potential bias $V$ applied across the array, see Fig.~\ref{fig:cctdiag}.  Each Josephson junction has a Josephson energy $E_J$, an effective capacitance $C_J$ and a capacitance to ground $C_G$.
The energy of the system is given by
\begin{equation}
H(\vec{q}\,) = \frac{1}{2}\vec{q\;}^T \mathbf{C}^{-1} \vec{q} + C_J V \vec{\delta}_{1}^T \mathbf{C}^{-1} \vec{q} 
\end{equation}
for a particular bias $V$ and charge configuration $\vec{q}=e\vec{n}$.  The source term $\vec{\delta}_1$ is equal to one for site $1$ and zero everywhere else. 

In contrast to previous theoretical work on JJ arrays~\cite{Homfeld:2011fw, ZorinAB:2006cv}, we take the conjugate phase to be across the ground capacitor $C_G$, which makes the charge variable $q_i$ simply the excess charge on the $i$th island, as is usually done when considering single-electron transistors.  Although this choice has no additional observable consequences, it simplifies the implementation of the kinetic Monte-Carlo algorithm for simulating the dynamics.

The capacitance matrix of the system is given by
\begin{equation}
\mathbf{C} = \left(  \begin{array}{cccc} 
	C_G + 2 C_J 	& -C_J	& 0	& \cdots \\
	-C_J	&	C_G + 2 C_J	& -C_J	& \cdots \\
	0	&	-C_J		&	C_G + 2 C_J	& \cdots \\
	\vdots	&	\vdots		&	\vdots		& \ddots
	\end{array} \right),
\end{equation}
where the inversion of this matrix can be computed very quickly numerically, but can also be expressed 
analytically~\cite{Huang:1997vn,Hu:1996uo,Hu:1996vj,Hu:1994vn,Hu:1995tv}, a fact we employ when deriving the interaction energy and threshold voltage.  For this circuit, we define an interaction length $\Lambda$ between charges in the array, where
\begin{equation}
\frac{1}{\Lambda} = \cosh^{-1}\left(1+\frac{C_G}{2 C_J} \right)
\end{equation}
and in the limit of small ground capacitance $\Lambda \approx \sqrt{C_J/C_G}$.

To understand the origin of correlated transport in a JJA, we first derive the effective interaction energy between charges within an array.  Taking the large array limit $N\gg\Lambda>1$, we find
\begin{equation}\label{eq:Un}
U(n_i,n_j) = \frac{1}{2 \sinh( \Lambda^{-1})}  \left[ \frac{n_i^2}{2 C_J} + \frac{n_j^2}{2 C_J}  + \frac{n_i n_j}{C_J} e^{-|i-j|/\Lambda} \right]
\end{equation}
which is the energy for two charges at positions $i$ and $j$ in an infinite length, zero-biased array.  In the limit of $C_G \ll C_J$, we can write $C_J \sinh \Lambda^{-1} \approx C_J/\Lambda \approx \sqrt{C_J C_G}$.  The first two terms in Eq.~(\ref{eq:Un}) are the respective charging energies and the third term is the interaction energy between the charges, which decays exponentially over the interaction length $\Lambda$.

We can also compute the threshold voltage for quasiparticles or Cooper-pairs to be injected from the voltage source,
\begin{equation}\label{eq:Vth}
V^{(\kappa)}_{\rm{th}} = \frac{\kappa e}{2 C_J [\exp(1/\Lambda)-1]}
\end{equation}
where $\kappa=1,2$ for quasiparticles or Cooper-pairs respectively.  Both the interaction energy and threshold voltage expressions are derived from pure energetic considerations.  The details of the array conductance and correlated transport depend critically on the form of the hopping rates for charges within the array, which in turn depend on the charge carrier and the influence of temperature, magnetic field and the specific noise model.

We are specifically interested in the small $E_J$ limit so we assume that coherent oscillations are completely suppressed and that the evolution of the system can be described using $P(E)$ theory\cite{Grabert:1991uv,Ingold:1992uo,Nazarov:2009wt}. Typically when computing the influence of noise on smaller circuits, the dominant source is the impedance of the leads and measurement circuitry.  In the case of JJ arrays however, this leads to the conclusion that noise dominates at the edges of the array but within the array its influence decays exponentially over the length scale $\Lambda$.  Although in principle this could lead to coherent behaviour deep within the array, we assume (as others have done~\cite{Ho:2010ft}) that within the array, locally generated noise dominates.  We specifically consider the noise generated by the internal impedance of the array itself, however additional contributions from background charge fluctuations~\cite{Gustafsson:2012tw} and two-state defects~\cite{Martinis:2005kw,Shalibo:2010ke,Schriefl:2006hw,Shnirman:2005go,Bushev:2010gw,Cole:2010kw} within both the junctions and the substrate may also play a role.  

To derive the internal impedance seen by a particular junction we assume that every other junction within the array has an impedance $Z_J$ given by a capacitor with capacitance $C_J$ and a resistor with resistance $R_T$ in parallel. Generally we assume that the resistance of each junction is large as compared to the resistance quantum, although the explicit value of $R_T$ proves to be unimportant. We neglect the Josephson inductance in the limit of small $E_J$. 

We can now formulate the effective circuit model (Fig.~\ref{fig:cctdiag}b) for the junction impedance $Z_J$ and the impedance of the rest of the array $Z(\omega)$.  Using the theory of continued fractions~\cite{Lorentzen:2008vx} we can derive the array impedance $Z(\omega)$ for an arbitrary length array, however the resulting expression is unwieldy.  To obtain a more useful expression, we derive an effective impedance by considering the resistive and capacitative response of the array individually.  Computing the resistance of half an array as $R_{HA} = N R_T/2$ and the effective half-array capacitance $C_{HA} = [C_G + \sqrt{C_G(C_G + 4 C_J)}]/2$,
we then approximate the total impedance of the half-array as that of the resistive and capacitative components separately (see Fig.~\ref{fig:cctdiag}c).
The total impedance seen by a junction is then given by $Z_t(\omega) = [i \omega C_J + Z(\omega)^{-1}/2]^{-1}$ where
\begin{equation}
Z(\omega)^{-1} = i \omega C_{HA} + \frac{2}{NR_T}
\end{equation}
is the contribution from each half-array (assuming the resistive and capacitive responses are decoupled).  In the large impedance regime we consider ($R_T\gg R_K=h/e^2$, $C_G \ll C_J$) this is a good approximation when compared to the (exact) continued fraction solution.

Integrating $\rm{Re}[Z_t(\omega)]$ over all frequencies, we obtain the amplitude for a delta-function approximation to the array impedance,
\begin{equation}
\rm{Re}[Z_t(\omega)] \approx \frac{2\sqrt{2}\pi\delta(\omega)}{\sqrt{(C_G+4 C_J)(C_G + 2 C_J + \sqrt{C_G(C_G+4 C_J)})}}
\end{equation}
which in the limit of $C_G \ll C_J$ gives $\rm{Re}[Z_t(\omega)] \approx (\pi/C_J)\delta(\omega)$ as expected~\cite{Ingold:1992uo}.
Approximating for short times, we can then evaluate the integral over the time correlation function~\cite{Ingold:1992uo} and obtain an expression for the $P(E)$ function, 
\begin{equation}
P_\kappa(E) = \frac{1}{\sqrt{4 \pi \kappa^2 E_c k_B T}}\exp\left[ \frac{-(E-\kappa^2 E_c)^2}{4 \kappa^2 E_c k_B T} \right]
\end{equation}
where the effective charging energy $E_c = e^2/2C_{A}$ is now that of the array impedance,
\begin{equation}
C_{A} = \frac{1}{2\sqrt{2}} \sqrt{(C_G+4 C_J)(C_G + 2 C_J + \sqrt{C_G(C_G+4 C_J)})}
\end{equation}
As this P(E) function describes the contribution due to noise generated within the array itself, this contributes to hopping rates for all the processes within the array.  In principle, close to the edges of the array, this noise contribution will be modified by the external impedance but for simplicity we ignore this extra contribution, given that the length of the array considered is ten times the interaction length.

To study the transport of (correlated) current in the superconducting limit, we must consider the contribution of both Cooper-pairs and equilibrium quasiparticles.  Although non-equilibrium quasiparticle distributions are possible and have been found to be an important contribution to loss processes in small JJ circuits~\cite{Martinis:2009bd,Saira:2012up,Leppakangas:2012uz}, we ignore these contributions as the assumption of thermal equilibrium is adequate when considering the transport through JJ arrays at moderate effective electron temperatures.  For similar reasons, we do not consider co-tunnelling processes in this analysis as the first order process dominates in all parameter regimes of interest.

The rate for an equilibrium quasiparticle to move between two charge states which differ by total energy $\delta E$ is given by
\begin{widetext}
\begin{eqnarray}\label{eq:EQP}
\hspace{-1.5cm}\Gamma_{\rm{eqp}}(\delta E) = \frac{1}{e^2R_T} \iint_{-\infty}^{\infty} d\epsilon \, d\epsilon' \, \frac{\mathcal{N}(\epsilon)}{\mathcal{N}(0)}\frac{\mathcal{N}(\epsilon' +\delta E)}{\mathcal{N}(0)} f(\epsilon) [1-f(\epsilon' + \delta E)] P_1(\epsilon-\epsilon'), \nonumber \\
\end{eqnarray}
\end{widetext}
where ${\mathcal{N}(\epsilon)}/{\mathcal{N}(0)}$ is the normalized BCS density of states, $f(\epsilon)$ is the Fermi function and the change in energy $\delta E$ also takes into account the chemical-potential shifts associated with charges entering or leaving the leads.  This rate scales linearly with $1/R_T$ above a characteristic gap whose size is set by both the superconducting gap $\Delta$ and the P(E) function which is centred around $E_c$.

In the limit of incoherent Cooper-pair transitions, the Cooper-pair hopping rate is simply given by
\begin{equation}
\Gamma_{\rm{cp}}(\delta E) = \frac{\pi}{2\hbar}E_J^2 P_2(\delta E). \label{eq:CP}
\end{equation}
The asymmetry of this function provides a compelling argument as to why the noise should be dominated by the large internal impedance of the array itself.  If one was to derive a hopping rate assuming vanishing impedance, this would result in a peak centred about zero and therefore approximately diffusive charge movement within the array.  Whereas in the large impedance limit, the lossy transport derives directly from the ease with which the array can absorb energy from the individual charges.  In addition, we find that the results obtained using the high-impedance environment are in better agreement with the experimentally observed current voltage characteristics~\cite{Bylander:2007gb, Schafer:2013ue}.

To simulate the time evolution of a JJA we use the kinetic Monte-Carlo (KMC) algorithm, whereby charge transitions are chosen stochastically based on the relative weight of the relevant transition rates~\cite{Bakhvalov:1989td,Voter:2007ul,Reuter:2011vq,Wasshuber:2001ta,Ho:2010ft}.  This results in an output consisting of successive charge configurations and the time between each transition, from which average quantities can be easily computed.  For all results we use a combination of initialisation and measurement phases in the Monte-Carlo simulation.  We initialise the system in the `empty' state, where there are zero charges on each island.  The system is then time-evolved according to the KMC algorithm for sufficient time that the charge distribution within the JJ array has equilibrated (typically $10^5$-$10^6$ Monte-Carlo steps).  The quantities of interest are then computed over a further $10^6$ time steps such that we observe variances of order $1\%$ for the average charge on a particular site within the 50 site array.

To study the correlated transport, we consider the charge-charge correlations within the array as this proves to be more computationally efficient than computing the current-current correlations directly.  
A key observable for Josephson junction arrays is the average charge on the $j$th site, $\langle n_j \rangle$.  Computing the charge-charge autocorrelation function $\langle n_j(\tau) n_j(0) \rangle$ on a particular site then gives a measure of the extent to which the charge distribution within the array is statistically correlated with itself at some later (or earlier) time $\tau$.  A further quantitive of interest is the Fourier transform of this correlation function, $\mathcal{F} [ \langle n_j(\tau) n_j(0) \rangle ]$, which gives a direct measure of the spectral response that would be observed in experiments~\cite{Bakhvalov:1989td,Bylander:2005gr,Bylander:2007gb,Bylander:2005gc}.
Although the Fourier transform of the charge-charge autocorrelation $\langle n_j(\tau) n_j(0) \rangle$ can be computed directly~\cite{Bakhvalov:1989td}, linearly sampling the resulting charge vector at a high bandwidth $>10$ GHz and then taking the autocorrelation and fast Fourier transform directly proves to be more efficient and less sensitive to numerical noise.

\section{Array conduction}


We are interested  in the conduction processes of arrays in the superconducting regime, with $E_J \ll E_c$ and non-negligible superconducting gap. This means that transport with both possible charge carriers, Cooper-pairs and quasiparticles, is strongly suppressed and therefore the experimental observation of not only conduction, but correlated conduction in this limit is somewhat surprising. In addition, we find theoretically that over a large range of temperatures and voltages, correlated conduction in such circuits is largely suppressed which implies a relatively small part of parameter space in which these correlations can be observed.  To focus on an experimentally relevant case, we consider a parameter regime (in terms of array length, $C_J$ and $C_G$) which corresponds approximately to that studied in Ref.~\onlinecite{Bylander:2007gb}. 

\begin{figure} [t!]
\centering{\includegraphics[width=0.9\columnwidth]{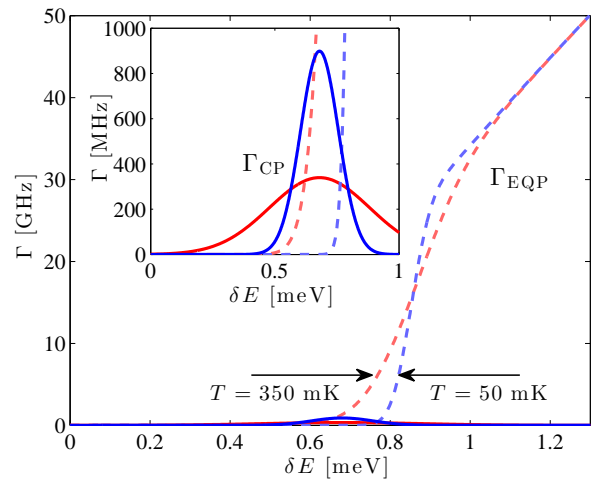}} 
\caption{Transition rate as a function of energy difference for both equilibrium quasiparticles (dashed red/blue) and incoherent Cooper-pairs (solid red/blue), with $E_J = E_c/20$.  The $P(E)$ peak (centred around $\delta E = 4 E_c$) and the superconducting step edge of the quasiparticle transition are broadened as a function of temperature.  Close to $\delta E = 2\Delta$ we see that the rate of quasiparticle and Cooper-pair transitions occur at similar rates.\label{fig:hoppingrates}}
\end{figure}

A key feature of the parameters chosen in our analysis is that the incoherent Cooper-pair rate has its maximum ($4E_c$) at the edge of the low temperature equilibrium quasiparticle gap ($2\Delta$). Therefore we choose the charging energy to be $E_c = 170$ $\mu$eV and the superconducting gap $\Delta = 340$ $\mu$eV, see Fig.~\ref{fig:hoppingrates}. In this regime, the position of the P(E)-function within the quasiparticle gap, means that incoherent Cooper-pair hopping cannot be considered a resonant process but in fact must compete directly with the \emph{equilibrium} quasiparticle rates, especially for large energy differences (larger than $\Delta$).  

Fig.~\ref{fig:hoppingrates} illustrates the hopping rates for quasiparticles and Cooper-pairs as a function of energy difference at an effective electron temperature of $T=50$~mK and $T=350$~mK, with $E_J = E_c/20 = 8.5$ $\mu$eV.  The effect of the $P(E)$ function is to broaden the Cooper-pair rates as well as increasing the gapped region in the quasiparticle rates and smoothing the edge of this gap.  The net result is that over a large range of energies both Cooper-pairs and quasiparticles can hop with approximately equal rates, especially at the typical transport voltages applied to this system. 

Throughout our analysis, we model an array of $N=50$ junctions with values of the junction capacitance $C_J=412.6$~aF and ground capacitance $C_G=25.9$~aF such that $\Lambda=4$. 
As we investigate the effect of varying $E_J$ while keeping the ratio $\Delta/E_c=2$, we consider junction resistances varying between $R_T=5 R_K = 129$~k$\Omega$ and $R_T=125 R_K = 3.23$~M$\Omega$ according to the standard relationship\cite{Tinkham:2004un}, 
\begin{equation}
E_J  = \left( \frac{\hbar}{2 e R_T} \right) \left( \frac{\pi \Delta}{2 e} \right) \tanh \left( \frac{\Delta}{2 k_B T} \right).
\end{equation}

\begin{figure} [t!]
\centering{\includegraphics[width=0.9\columnwidth]{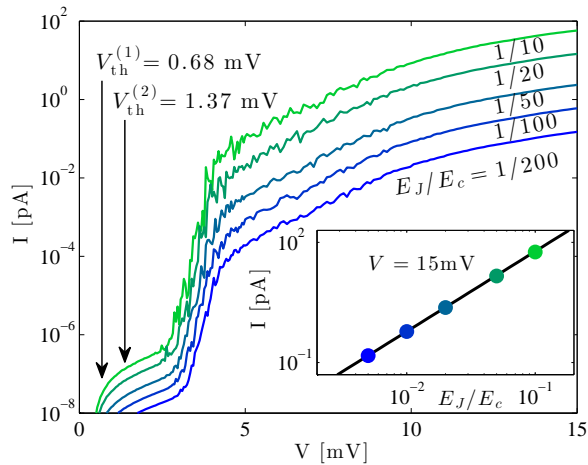}} 
\caption{Current-voltage characteristics as a function of $E_J$ (logarithmic current scale) at an effective electron temperature of $T=200$mK.  The dominant effect of varying the Josephson energy is to scale the current proportional to $E_J^2$.  The threshold voltages for injection of quasiparticles and Cooper-pairs, Eq.~\ref{eq:Vth}, is illustrated with arrows whereas a further transition to a high current state occurs at higher voltages ($V\approx 3-4$ mV).  Insert: At fixed $V=15$ mV, the current scales as $E_J^2$, indicated by the black line.\label{fig:IVcurveEJ}}
\end{figure}

The absolute magnitude of the current-voltage (I-V) characteristics of the array depends on both $E_J$ and temperature.  For our chosen parameters we find that conduction is sporadic with a large variance for electron temperatures less than $T\approx 150$mK as the system can become trapped in meta-stable charge states which do not easily decay due to the large quasiparticle gap.  These trapping states are inconsistent with the observed reproducible and smooth response seen in experiments~\cite{Bylander:2007gb} and their absence implies an effective electron temperature which is higher than the nominal base temperature of the experiments.  

A fundamental assumption of $P(E)$ theory is that the energy associated with hopping of charges is dissipated within the circuit itself and in the quasiparticle degrees of freedom.  It is this excess energy which can lead to higher effective electron temperatures, an effect which has been observed in qubit~\cite{Martinis:2009bd,Barends:2011eh,Manninen:2012ti} and SET experiments~\cite{Gustafsson:2012tw}.  The effective temperature seen by the charge carriers can depend in general on the bias conditions, junction properties, the superconducting gap and even the timescale over which the experiment is conducted~\cite{Gustafsson:2012tw, Maisi:2013vu, Heimes:2014fc}.  For simplicity in the remainder of this analysis, we assume an electron temperature of $T=200$~mK.

Fig.~\ref{fig:IVcurveEJ} shows the I-V response at $T=200$~mK as a function of Josephson energy, demonstrating a characteristic $E_J^2$ dependence (see insert to Fig.~\ref{fig:IVcurveEJ}).  This indicates that the dominant processes are Cooper-pair hopping events, Eq.~(\ref{eq:CP}).  More interesting is the behaviour of the threshold for conduction, which goes through two distinct transitions.  Initially conduction rises rapidly once the voltage is greater than the threshold for injection of quasiparticles is reached (although there is some thermal broadening of this threshold).  At this point the simulations clearly predict that conduction is possible, however it is at extremely low currents which are well below the detection threshold of existing experiments.  A second distinct transition to observable current levels is observed in the range $3-4$~mV, which one could associate with the injection of Cooper-pairs.  This however is not the case as the injection threshold for Cooper-pairs is only a factor of two larger than that for quasiparticles, which is well below the observed threshold.  In fact, this second transition to a higher current state is associated with the interplay of quasiparticles and Cooper-pairs and the `filling' of the array, as will become apparent in section 5.

\section{Correlated transport within the array: low voltages}

\begin{figure} [t!]
\centering{\includegraphics[width=0.95\columnwidth]{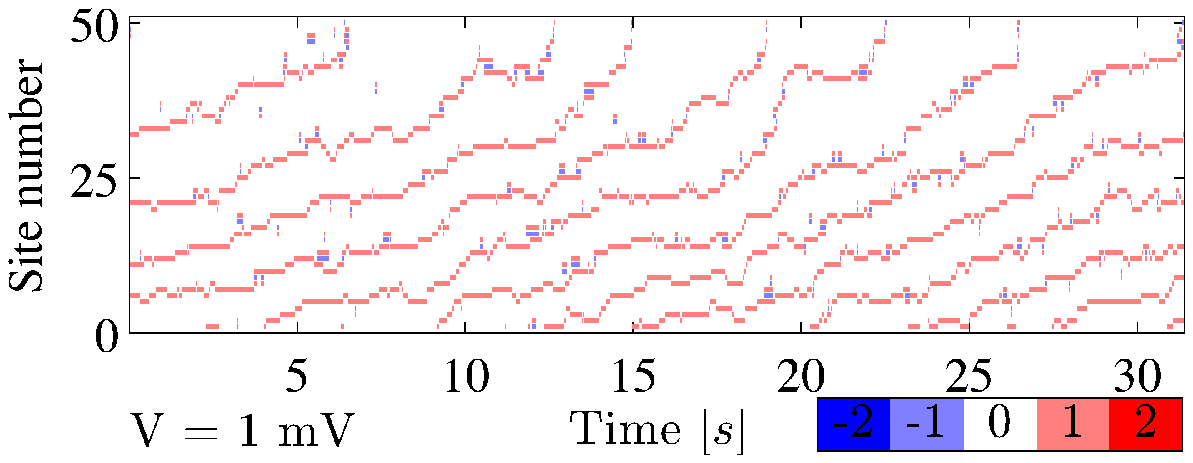}}  \\
\centering{\includegraphics[width=0.95\columnwidth]{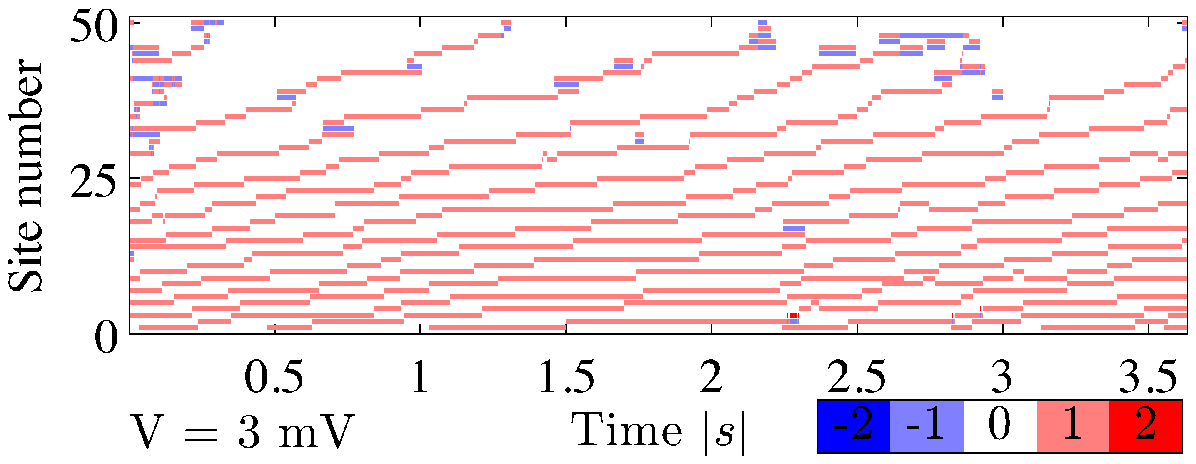}} \\
\vspace{0.1cm}
\centering{\includegraphics[trim=0 5 0 10, clip, width=0.4\columnwidth]{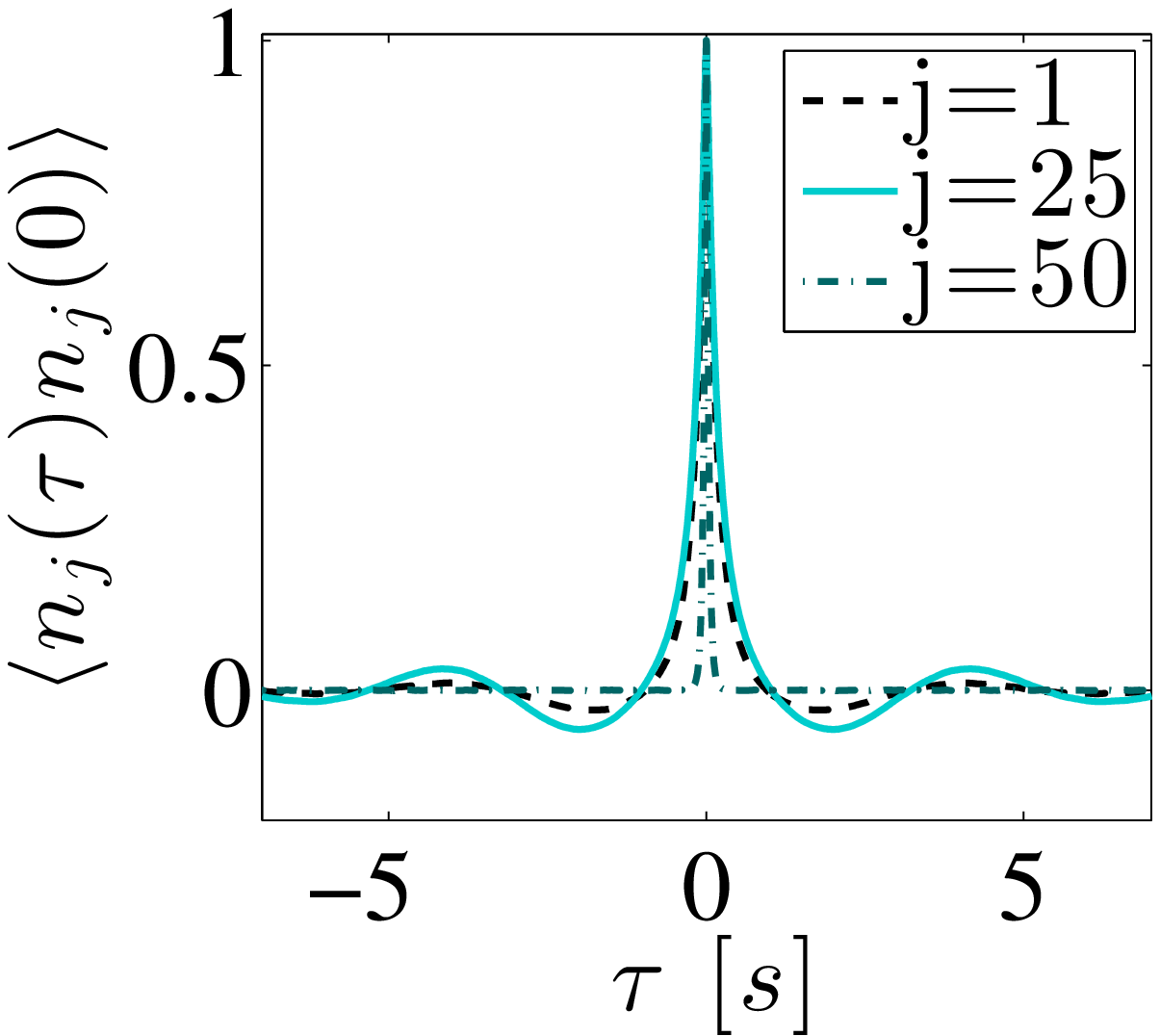}} \quad \centering{\includegraphics[trim=0 5 0 10, clip, width=0.4\columnwidth]{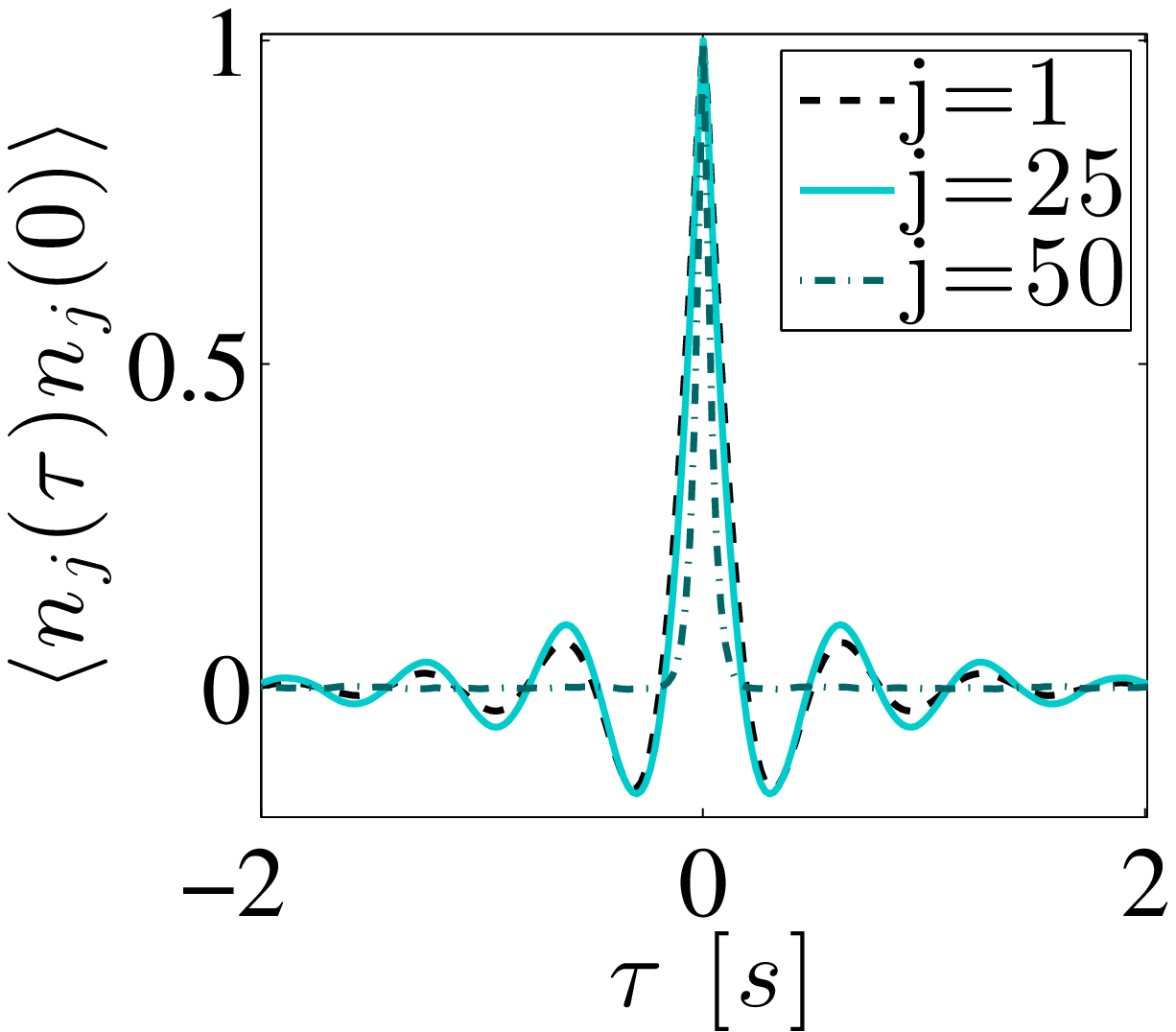}} \\
\rput(-3.8cm,10.1cm){(a)}
\rput(-3.8cm,6.6cm){(b)}
\rput(-3.8cm,3.3cm){(c)}
\rput(0.1cm,3.3cm){(d)}
\caption{Charge distribution within the array for an example Monte-Carlo instance, as a function of time at two different bias voltages; a) $V=1$ mV, b) $V=3$ mV.  Just above the conduction threshold ($V=1$~mV) we see diffusive type transport dominated by the movement of quasiparticles.  The charge distribution within the array is quasi-periodic (correlated) but the movement of charge is partially diffusive.  As the voltage increases, the correlations become stronger due to higher charge densities and less diffusive `jitter' of the charge carriers.  The charge occupancy autocorrelation function $\langle n_j(\tau) n_j(0) \rangle$ is plotted in c) $V=1$ mV and d) $V=3$ mV for 3 different sites, ie.\ the beginning, middle and end of the array.  We see stronger temporal correlations in the middle of the array than at the beginning, whereas at the end we see almost no correlations.  The increased oscillation amplitude at $V=3$ mV also illustrates the increase in correlation strength at voltages and charge densities.\label{fig:sta}}
\end{figure}

We now turn our attention to the correlated transport of charges through the array.  To observe such correlations, the low voltage (close to threshold) regime is usually considered, such that the average charge density is low and the charges are separated by approximately $\Lambda$, see for example Refs.~\onlinecite{Bylander:2005gc, Walker:2013wp}.  This conventional correlated transport regime is illustrated in Fig.~\ref{fig:sta}a) and b), where we plot the charge distribution within the array as a function of time for two different voltages within the `low current' state.  At very low voltage (just above threshold), the conduction is partially diffusive due to the relative high electron temperature ($T=200$~mK).  As the voltage increases ($1$ mV$\lesssim V \lesssim 4$ mV), the charge distributions become progressively more correlated and more closely mimics the `conventional' correlated transport in normal arrays~\cite{Walker:2013wp}. We see a characteristic periodic distribution of charges within the array a fixed point in time, due to the interplay between the applied voltage and the repulsive interaction between charges.  As the charges move systematically along the array, this spatial correlation between the charges also manifests as a temporal correlation of the type observed in experiments~\cite{Bylander:2005gc}.

To more clearly see this temporal correlation between the charges, in Fig.~\ref{fig:sta}c) and d) we show the charge occupancy autocorrelation function as a function of delay time $\tau$.  The damped oscillations illustrate that the charge occupancy on a particular site is correlated with occupancy at some characteristic time later.  This characteristic time corresponds to the mean spacing between charges as they travel along the array.  We see that the amplitude of the oscillations is stronger at large voltage biases as the charges are packed more tightly together, reducing the thermal `jitter' observed just above threshold.  

This temporal correlation also varies with position within the array, being strongest in the middle, weaker at the start (as the voltage source is uncorrelated) and non-existent at the end of the array (for this voltage bias configuration).  This variation in correlation strength has been studied previous for different bias configurations~\cite{Walker:2013wp} and arises due to two effects.  First and foremost, for this bias configuration the average charge distribution is approximately linear across the array, approaching zero at the end which is connected to ground.  As each charge spends a vanishing amount of time on the final site, the corresponding autocorrelation function does not show the temporal correlations.  There is also a more subtle effect which arises at low voltage bias, due to the periodic distribution of charges being partially `pinned' by the boundary conditions in the array.  This results in a slight but periodic modulation of the \emph{average} charge distribution.  These effects have been studied in depth for the case of normal conduction in Ref.~\onlinecite{Walker:2013wp}.  In this analysis, we are specifically interested in the interplay of normal and superconducting processes and therefore limit ourselves to the charge occupancy statistics measured at a fixed point in the middle of the array.

\begin{figure} [t!]
\centering{\includegraphics[width=0.9\columnwidth]{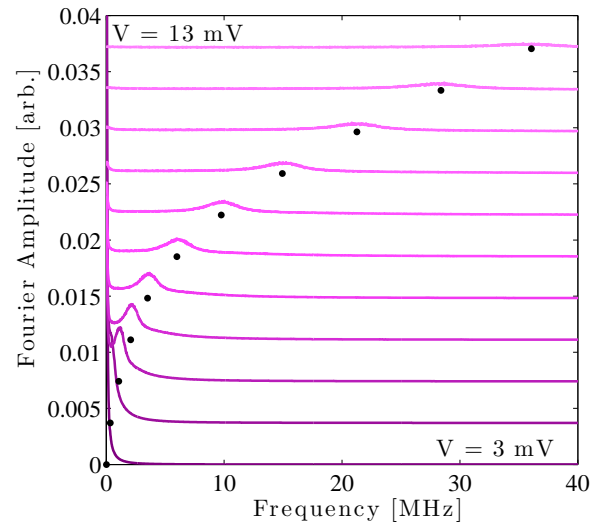}} 
\caption{Spectral response of the charge-charge autocorrelation function measured in the centre of the array, for increasing voltage in steps of $1$~mV (curves vertically displaced for clarity).  As the voltage increases, the peak frequency increases linearly with increasing current.  The strength of the correlations are reduced at higher voltages as these correspond to higher average charge densities, reducing the formation of quasi-periodic charge states~\cite{Bakhvalov:1989td,Bylander:2005gc,Walker:2013wp}.  The dots indicate the position of the peak frequency computed from the magnitude of the current via $f_p = I/Q$ where $Q=2$ for Cooper-pairs, indicating Cooper-pair dominated correlated transport.\label{fig:correlationpeaks}}
\end{figure}

\section{Correlated transport within the array: high voltages}

\begin{figure*} [t!]
\centering{\includegraphics[trim=5 0 25 0, clip, width=\textwidth]{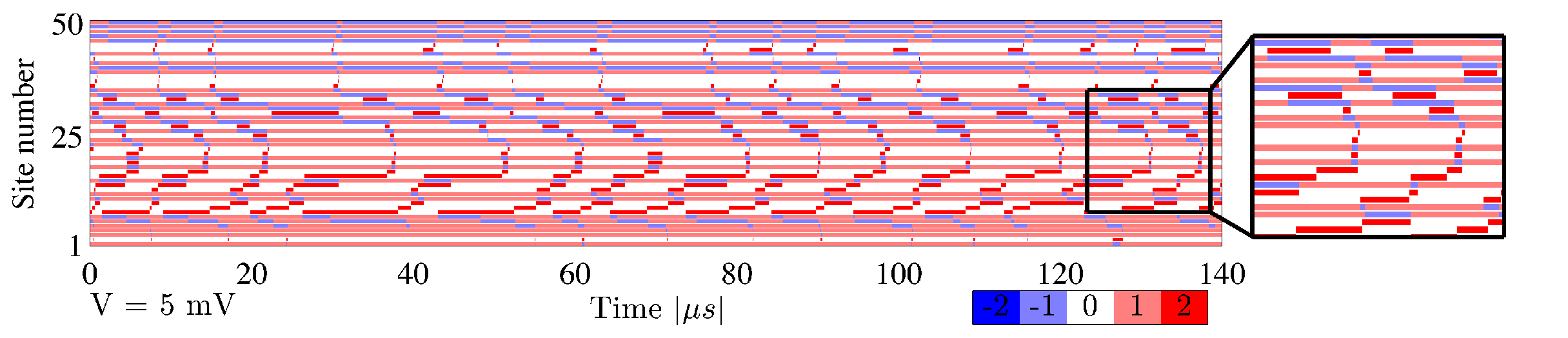}}
\caption{Charge distribution within the array for an example Monte-Carlo instance, as a function of time at $V=5$~mV. At higher voltages (compared to Fig.~\ref{fig:sta}) a new mechanism for correlated transport forms, consisting of a quasi-static `quasiparticle gas' through which transport is carrier by movement of Cooper-pair states.  Conduction proceeds via interconversion of even charges states ($n=+2$) moving forward through the array and odd charge states ($n=-1$) moving backwards. As well as being strongly correlated, this form of conduction proves to be more stable to background disorder and results in a current several orders of magnitude larger than would be otherwise observed.\label{fig:spacetimeb}}
\end{figure*}

To investigate correlated transport at higher voltages, we plot the spectral response (Fourier transform of the charge-charge autocorrelation function) measured in the middle of the array, see Fig.~\ref{fig:correlationpeaks}.  The peaks, corresponding to temporally correlated charge states, form almost immediately once the circuit transitions to the higher current state and are still observable up to at least $V=13$~mV.  The average charge density at these voltages is $\langle n \rangle \approx 1.5$ which is well above the regime usually considered.  For comparison with earlier work~\cite{Walker:2013wp}, in this model the charge distribution has already reached a linear drop throughout the array well before the transition to high current.  

At these higher voltages, the correlations derive from a more complicated mechanism involving both single-charge (quasiparticle) and double-charge (Cooper-pair) excitations, which is not seen in a normal conducting array (without superconductivity).  This additional transport mechanism is responsible for the transition to a higher current state for $V\gtrsim5$ mV (see Fig.~\ref{fig:IVcurveEJ}). Here the voltage is large enough to push the charge excitations closer together so that the interplay between quasiparticle and Cooper-pair tunneling, with the corresponding island charges, is energetically favourable and charges can flow freely again through the array.  

\begin{figure*} [t!]
\centering{\includegraphics[width=0.9\textwidth]{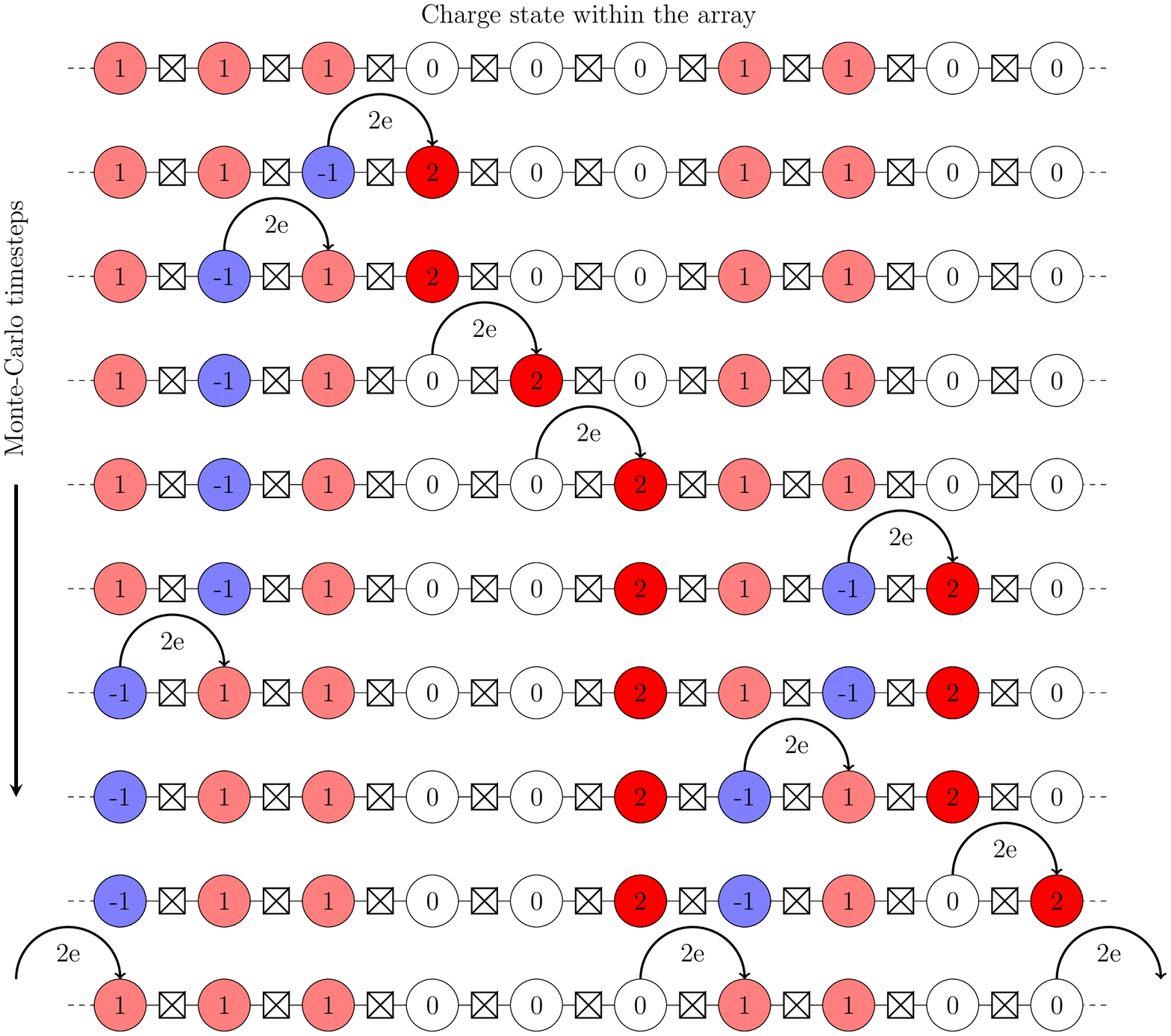}}
\caption{Diagram illustrating the transport mechanism for charges in the low magnetic field, high voltage limit.  A quasi-static background of singly occupied states forms through the array.  At the boundaries of these regions (where two neighbouring sites are $n=1$ and $n=0$ respectively) a dipole state can be created which consists of neighbouring $n=-1$ and $n=2$ charge states.  This dipole state can then immediately separate which results in a net current as the $+2$ excitation moves right and the $-1$ excitation moves left.  All charge movement events correspond to a charge of $2e$ moving left to right, ie.\ via incoherent Cooper-pair hopping.  The creation and annihilation of dipole states in this way allows Cooper-pair dominated transport to dominate, even in the presence of a static background of singly occupied ($+1$) states.\label{fig:chargehoppingdiagram}}
\end{figure*}

The exact mechanism driving this interplay of odd and even charge states can be seen in Fig.~\ref{fig:spacetimeb} and~\ref{fig:chargehoppingdiagram}.  Odd charge states ($n=1$) form which are relatively stationary, on the time scale of the current flow, and can sit on neighbouring sites due to the high bias voltage.  These quasiparticle states form a quasi-static background charge distribution throughout the array.  Transport then proceeds via movement of incoherent Cooper-pairs where the moving charge states are either $n=+2$ when there is a background charge of $n=0$, or $n=-1$ when the background state is $n=+1$.  Fig.~\ref{fig:chargehoppingdiagram} illustrates the conduction pathway for this transport mechanism where the charge distribution across 10 sites within the array is depicted for a series of example Monte-Carlo steps.  Although we see a combination of charge states $n=-1,0,1$ and $2$, current is always carried by movement of Cooper-pairs from left to right.  

The stability of this transport mechanisms relies on the high average charge density and the stationary quasiparticle states (due to the gapped hopping rate, see Fig.~\ref{fig:hoppingrates}).  It is only when both conditions are met, as a result of both high voltage and vanishing magnetic field, that we see this transport mechanism, where incoherent Cooper-pair hopping accounts all of the current flow through the array, once the system has equilibrated.

This new transport mechanism has a lower effective resistance as it is dominated by the Cooper-pair hopping rate $\Gamma_{\rm{CP}} \propto E_J^2$ rather than simply the normal resistance $\Gamma_{\rm{CP}} \propto 1/R_T$, resulting in a much higher current state than is seen at lower voltages.  The interplay between $n=+2$ states and $n=-1$ states means that periodic spacing between current carriers need only be maintained for a few sites before conversion between charge states, which results in a more even distribution of correlations throughout the array.  For even higher voltages ($V\gg10$~mV), this effect is again washed out due to higher occupancy charge states and the correlations disappear.  

This new transport regime also proves to be more robust in the presence of disorder, particularly in comparison to the zero-bias conductance regime~\cite{Zimmer:2013dg}.  Performing the same simulations with background disorder modelled as an initial random fractional charge state~\cite{Johansson:2001tc} allows us to mimic how the system would respond to random static variations in the local electric potential on each site.  We initialise the system with charge states $n\in (-1,1)$, we then evolve the system for enough time to reach equilibrium before collecting statistics.  As the transport mechanisms considered only move charge from one site to another, the fractional offset due to the initial disorder is preserved modulo $1e$.  

This method of modelling disorder results in no significant change in the I-V characteristics or correlation response in the high current regime.  This is because the high current regime relies on the formation of a random distribution of static `background' charges, the exact distribution of which varies from run to run.  Including disorder simply selects one particular distribution (or a subset thereof) preferentially to minimise the total energy of the system.  The transport mechanism proceeds in exactly the same fashion as without disorder, via incoherent Cooper-pair hopping through the background charge distribution.  It should be noted that disorder does have a significant effect on the initial (quasiparticle dominated) low voltage regime, although as stated earlier, this regime is not experimentally resolvable for these parameters.  

\section{Magnetic field dependence of correlated transport}

\begin{figure} [t!]
\centering{\includegraphics[width=0.8\columnwidth]{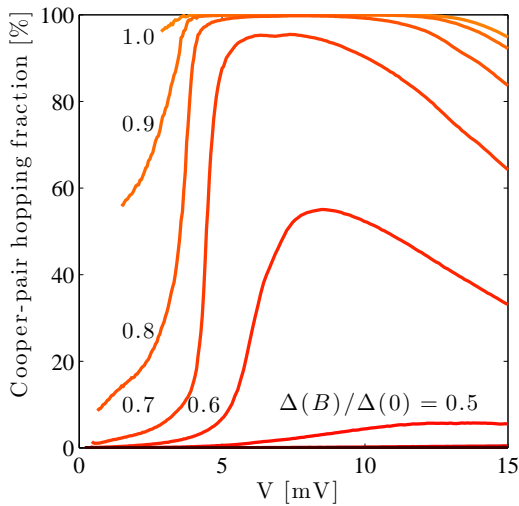}} 
\caption{The relative fraction of Cooper-pair hopping as a percentage of the total hopping events within the array, as a function of both voltage and magnetic field (plotted for $I>0.5$ aA).  At zero magnetic field ($\Delta(0) = 340$ $\mu$eV), the transport is almost entirely dominated by incoherent Cooper-pair hopping.  As the magnetic field is increased ($\Delta(B) \rightarrow 0$) the role of Cooper-pair hopping decreases and quasiparticles dominate.\label{fig:fracCP}}
\end{figure}

A key motivation for this analysis is the experimentally observed correlated transport as a function of magnetic field.  At this point we have demonstrated that correlated transport can occur in the zero magnetic field, small $E_J$ limit.  We now consider the characteristics of the current flow as a function of an applied magnetic field.  We assume the Josephson energy displays the usual magnetic field dependence~\cite{Tinkham:2004un}, 
\begin{equation}
E_J(B) =   \left( \frac{\hbar}{2 e R_T} \right) \frac{\pi\Delta(B)}{2e} \tanh\left( \frac{\Delta(B)}{2 k_B T} \right),
\end{equation}
therefore both the superconducting gap and $E_J$ will contribute a magnetic field dependence to the conduction properties.  As the magnetic field increases, the closing of the superconducting gap increases the rate of quasiparticle hopping while the Cooper-pair rate is dropping due to the reduction in $E_J$.  For our parameters, this results in non-monotonic variation in the current for given voltage and the disappearance of the second (high current) transition for $\Delta(B)/\Delta(0) \lesssim 0.5$.  In Fig.~\ref{fig:fracCP} we plot the relative fraction of Cooper-pair hopping compared to all hopping events as a function of both voltage and magnetic field.  To avoid numerical noise from the finite simulation time, we only include data for currents greater than $0.5$ aA.  Looking at the relative contribution of Cooper-pair to quasiparticle events, we see that for $\Delta(B)/\Delta(0) \gtrsim 0.5$ Cooper-pairs dominate in the region of interest ($4$ mV$ \lesssim V \lesssim 15$ mV).  For lower voltages, quasiparticle transport plays a more important role as the charge density within the array is closer to zero.  Quasiparticles also dominate at large magnetic fields (small superconducting gaps) as one would expect.

\begin{figure} [t!]
\centering{\includegraphics[width=0.9\columnwidth]{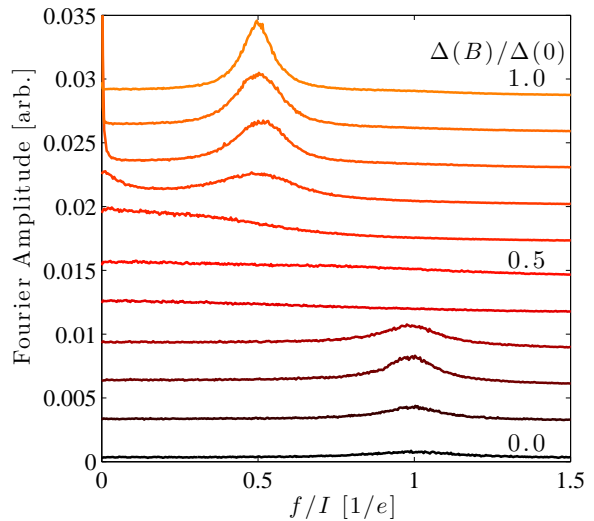}} 
\caption{Spectral response as a function of relative gap size for $T=200$ mK and $E_J=E_c/20$ (curves vertically displaced for clarity). The frequency axis is normalised $f/I$ such that for $f/I=1/Q = 1/e$ for quasiparticle transport and $f/I=1/Q = 1/2e$ for Cooper-pair transport, ie.\ $Q$ defines the magnitude of the effective charge carrier.  At small gap sizes (high magnetic fields) we see quasiparticle dominated transport, whereas at large gap sizes (low magnetic field) we observed Cooper-pair dominated transport plus an additional zero frequency component, related to slow rearrangement of the background charge.\label{fig:VaryGapSpectra}}
\end{figure}

We now consider the spectral response of the charge-charge correlations at fixed current ($I=1$ pA) as a function of superconducting gap size.  Fig.~\ref{fig:VaryGapSpectra} illustrates the spectral response as a function of gap size where correlated transport is observed in both the small and large gap limits, although the correlations disappear in the intermediate regime.  The frequencies where we observe a peak in the spectral response in the low and high gap regimes differ by a factor of two, due to the dominant charge carriers being either quasiparticles ($Q=1e$) or Cooper-pairs ($Q=2e$) where $Q$ defines the magnitude of the effective charge carrier.  This is consistent with magnetic field dependence of the relative fraction of Cooper-pairs plotted in Fig.~\ref{fig:fracCP} as well as the experimentally observed relationship between frequency response and current in Ref.~\onlinecite{Bylander:2007gb}.

Although the distribution of single occupancy excitations in the high current regime is approximately static on the time scale of the current flow, long time fluctuations are observed which correspond to random rearrangement of the quasiparticle background.  These slow fluctuations can be seen as fluctuations of the current response in the high current transition regime in Fig.~\ref{fig:IVcurveEJ}.  This slow background charge rearrangement leads to a non-negligible zero frequency component at low magnetic field, see Fig.~\ref{fig:VaryGapSpectra}.  As the magnetic field is increased, this contribution to the spectra broadens as the hopping rate for quasiparticles increases.  This transition regime corresponds to where the high current transport regime is no longer correlated due to strong fluctuations of the quasi-static background but the overall charge density is still too high to observe conventional (normal) correlated conduction.  The exact details of this transition depend strongly on the characteristics of the array (effective impedance, temperature, Josephson energy).  At sufficiently high magnetic field, conduction is simply dominated by quasiparticles and we see a single frequency peak corresponding to correlated transport. 

This model reproduces the observation of correlated transport in both the high and low magnetic field limits~\cite{Bylander:2007gb}, however several important questions remain.  A key experimental observation is the continuous transition from $Q=1e$ to $Q=2e$ which is not reproduced in our simulations.  Although broadening due to the finite response of the rf-SET would explain the continuous transition, this would still require coexistence of both the $Q=1e$ to $Q=2e$ dominated phases which is not observed here.  However, given that both background disorder and the bias position of the rf-SET itself can stabilise particular quasi-static charge configurations, a plausible explanation is that the system fluctuates between both transport states, the result of which is then time averaged during rf-SET detection.  Secondly, we have not considered the effects of variations in the effective electron temperature and in the impedance seen by the charges as a function of time and/or position within the array.

\section{Conclusion}

We have developed a model for transport due to both incoherent Cooper-pairs and equilibrium quasiparticles which explicitly includes the ability of the array impedance itself to dissipate energy.  Such a model fundamentally modifies the response of the charge carriers and leads to qualitatively different predictions for the array response which help to explain several experimentally observed effects.  We observe correlated transport due to incoherent Cooper-pair hopping which can occur at higher temperatures and higher voltages than one would expect for normal (correlated) conduction.  This new correlated transport phase is carried via movement of double excitations (Cooper-pairs) through a random distribution of stationary single excitations (quasiparticles).  The breakdown of this phase as a function of applied magnetic field is consistent with the observation of correlated transport in both the low and high field regimes, even for small Josephson energy.

\section{Acknowledgements}
The authors wish to acknowledge useful discussions with A. Shnirman, T. Duty, P. Delsing, K. Walker and N. Vogt.  This work was supported by the Victorian Partnership for Advanced Computing (VPAC).
	
\bibliographystyle{apsrev4-1}
\bibliography{JJAsmallEJ}

\end{document}